\documentstyle[galley,epsf]{mn}

\newif\ifAMStwofonts
  \def\simge{\mathrel{\raise1.16pt\hbox{$>$}\kern-7.0pt
    \lower3.06pt\hbox{{$\scriptstyle \sim$}}}}           
  \def\simle{\mathrel{\raise1.16pt\hbox{$<$}\kern-7.0pt
    \lower3.06pt\hbox{{$\scriptstyle \sim$}}}}           
\ifoldfss
  \ifCUPmtlplainloaded \else
    \NewTextAlphabet{textbfit} {cmbxti10} {}
    \NewTextAlphabet{textbfss} {cmssbx10} {}
    \NewMathAlphabet{mathbfit} {cmbxti10} {} 
    \NewMathAlphabet{mathbfss} {cmssbx10} {} 
  \fi
  \ifAMStwofonts
    \ifCUPmtlplainloaded \else
      \NewSymbolFont{upmath} {eurm10}
      \NewSymbolFont{AMSa} {msam10}
      \NewMathSymbol{\upi}     {0}{upmath}{19}
      \NewMathSymbol{\umu}     {0}{upmath}{16}
      \NewMathSymbol{\upartial}{0}{upmath}{40}
      \NewMathSymbol{\leqslant}{3}{AMSa}{36}
      \NewMathSymbol{\geqslant}{3}{AMSa}{3E}

      \let\geq=\geqslant 
    \fi
  \fi
\fi 

\ifnfssone
  \newmathalphabet{\mathit}
  \addtoversion{normal}{\mathit}{cmr}{m}{it}
  \addtoversion{bold}{\mathit}{cmr}{bx}{it}
  \newmathalphabet{\mathbfit} 
  \addtoversion{normal}{\mathbfit}{cmr}{bx}{it}
  \addtoversion{bold}{\mathbfit}{cmr}{bx}{it}
  \newmathalphabet{\mathbfss} 
  \addtoversion{normal}{\mathbfss}{cmss}{bx}{n}
  \addtoversion{bold}{\mathbfss}{cmss}{bx}{n}
  \ifAMStwofonts
    \ifCUPmtlplainloaded \else
      \UseAMStwoboldmath
      \makeatletter
      \new@mathgroup\upmath@group
      \define@mathgroup\mv@normal\upmath@group{eur}{m}{n}
      \define@mathgroup\mv@bold\upmath@group{eur}{b}{n}
      \edef\UPM{\hexnumber\upmath@group}
      \new@mathgroup\amsa@group
      \define@mathgroup\mv@normal\amsa@group{msa}{m}{n}
      \define@mathgroup\mv@bold\amsa@group{msa}{m}{n}
      \edef\AMSa{\hexnumber\amsa@group}
      \makeatother
      \mathchardef\upi="0\UPM19
      \mathchardef\umu="0\UPM16
      \mathchardef\upartial="0\UPM40
      \mathchardef\leqslant="3\AMSa36
      \mathchardef\geqslant="3\AMSa3E

      \let\geq=\geqslant 
    \fi
  \fi
\fi 

\ifnfsstwo
  \DeclareMathAlphabet{\mathbfit}{OT1}{cmr}{bx}{it}
  \SetMathAlphabet\mathbfit{bold}{OT1}{cmr}{bx}{it}
  \DeclareMathAlphabet{\mathbfss}{OT1}{cmss}{bx}{n}
  \SetMathAlphabet\mathbfss{bold}{OT1}{cmss}{bx}{n}
  \ifAMStwofonts
    \ifCUPmtlplainloaded \else
      \DeclareSymbolFont{UPM}{U}{eur}{m}{n}
      \SetSymbolFont{UPM}{bold}{U}{eur}{b}{n}
      \DeclareSymbolFont{AMSa}{U}{msa}{m}{n}
      \DeclareMathSymbol{\upi}{0}{UPM}{"19}
      \DeclareMathSymbol{\umu}{0}{UPM}{"16}
      \DeclareMathSymbol{\upartial}{0}{UPM}{"40}
      \DeclareMathSymbol{\leqslant}{3}{AMSa}{"36}
      \DeclareMathSymbol{\geqslant}{3}{AMSa}{"3E}

      \let\geq=\geqslant 
    \fi
  \fi
\fi 

\ifCUPmtlplainloaded \else
  \ifAMStwofonts \else 
    \def\upi{\pi}
    \def\umu{\mu}
    \def\upartial{\partial}
  \fi
\fi

\title{Turbulent Compressibilty of Protogalactic Gas}
\author[John Scalo and Anirban Biswas]
	{John Scalo and Anirban Biswas \\
	 Department of Astronomy, University of Texas, Austin, TX 78712-1083,USA}

\pagerange{\pageref{firstpage}--\pageref{lastpage}}

\begin{document}

\maketitle

\label{firstpage}

\begin{abstract}
The star formation rate in galaxies should be related to the
fraction of gas that can attain densities large enough for
gravitational collapse.  In galaxies with a turbulent interstellar medium, this
fraction is controlled by the effective barotropic index $\gamma = {\rm dlog P/dlog}{ \rho}$
 which measures the turbulent compressibility.  When the
cooling timescale is smaller than the dynamical timescale, $\gamma$ can be
evaluated from the derivatives of cooling and heating functions, using the
condition of thermal equilibrium.  We present calculations of $\gamma$ for
protogalaxies in which the metal
abundance is so small that ${\rm H_2}$ and HD cooling dominates.  For a
heating rate independent of temperature and proportional to the first
power of density, the turbulent gas is relatively ``hard", with $\gamma \ga   1$, at large densities, but
 moderately ``soft", $\gamma \la 0.8$, at densities
below around $10^4 {\rm cm}^{-3}$. At low temperatures the density probability distribution should fall rapidly for densities larger than this value,
 which corresponds physically to the critical density at which collisional and radiative deexcitation rates of HD are equal. 
The densities attained in turbulent protogalaxies thus depend on the relatively large deuterium abundance in our universe. 
We expect the same physical effect to occur in higher metallicity gas with different coolants. The case in which adiabatic (compressional) heating due to cloud collapse dominates is also discussed, and suggests a criterion for the maximum mass of Population III stars.
\end{abstract}

\section{ Introduction }

A physical understanding of the star formation rate (SFR) in galaxies
remains elusive.  Probably the only consensus is that the SFR must be
controlled by the rate of formation of substructure, or ``clouds", dense
enough that they can undergo gravitational collapse and fragmentation. 
Whether the rate of formation of dense clouds is controlled by
large-scale gravitational instabilities, thermal instability, turbulent
compression, or some other process is currently unknown, although all
these possibilities have been discussed in various degrees of detail.  

Of particular interest is the SFR in high-redshift galaxies.  The
reionization of the intergalactic medium and the existence of heavy
elements and dust in galaxies with redshifts up to ${\rm z} \sim 5$ (see discussion in sec 4.2 below) suggest that the first luminous objects must have appeared at
larger redshifts.  A large number of papers have tried to estimate the
SFR and the mass of these early objects, the latter based mostly on a criterion that
the cooling time due to ${\rm H_2}$ rovibrational transitions be smaller than the
collapse time (e.g. Ostriker \& Gnedin 1996, Tegmark et al. 1997, Abel et
al. 1998, Nishi  \& Susa 1999, and many earlier references given in these
papers).  The problem is complicated because the ${\rm H_2}$ abundance is
generally not in equilibrium (Shapiro \& Kang 1987, Galli \& Palla 1998 and
references therein) and because there is feedback between the UV
background generated by the earliest stars and the ${\rm H_2}$ abundance, a
feedback which may be negative (e.g. Haiman, Abel \& Rees 1999) or
positive (Ferrara 1998).  The chemistry, cooling, hydrodynamics, and
radiative transfer are all complex (e.g. Abel et al. 2000 and references therein).  Norman \& 
Spaans (1997) and Spaans \& Norman (1997) proposed, using a very detailed
model for the chemistry, cooling, and radiative transfer (but not the SFR
or hydrodynamics), a scenario in which the SFR is kept at low levels
until fine structure cooling by metals allows a thermal instability,
resulting in the enhanced formation of dense clouds.

    What all these studies have in common is that they consider the gas
to be quasi-static, except for gravitational collapse.  On the other
hand, there is overwhelming evidence that the Milky Way and nearby
galaxies are turbulent in some sense, with supersonic motions occurring
over a very large range of scales (see the papers in Franco \& Caraminana 1999 for a survey).
  Even in protogalaxies for which there
is no stellar input to drive the turbulence, it seems plausible that the
gas will be turbulent.  The Reynolds numbers are extremely large, so
shear and compression can be amplified and transferred to smaller or larger scales  nonlinearly by fluid advection,
and there is probably no magnetic field (or at least only a weak one)
which might suppress the motions.  The gravitational potential of the
inhomogeneous protogalaxy may be the original energy source.  When one
considers that it is not easy to separate monolithic collapse from the
accretion of smaller, perhaps transient, gas clouds, the protogalactic ``turbulence"
might resemble smaller-scale versions of the dynamically and spatially irregular systems modeled by Haenelt et al. (1998). However,
numerical simulations are incapable of determining whether small-scale turbulence
will occur because of lack of spatial resolution.\footnote{It should be
realized that even terrestrial turbulence is simply a ubiquitous
empirical fact which numerical simulations attempt to model, but could
not predict, at least until very recently.} 
Some weak evidence that protogalaxies
may be turbulent was also found in the simulations of Kulsrud et al. (1997). 
Turbulence in protogalaxies might also alleviate the long-standing
``cooling problem" for protodisk systems (see Navarro \& Steinmetz 1997 and references therein)
by reducing the efficiency of star formation even in the presence of
strong cooling.

It is therefore of interest to examine the ability of turbulent
motions to generate large density contrast clouds in protogalaxies 
with very small metal abundances.
   Compressible turbulence causes the gas in
galaxies to ``slosh" around, giving rise to a spectrum of density
fluctuations. Simulations and phenomenological studies (Scalo et al. 1998, Passot \& Vazquez-Semadeni 1998) for turbulent galaxies have shown that the density probability distribution (density) function ${\rm f}(\rho) $   may be controlled by the turbulent compressiblity or the effective barotropic index $\gamma = d \log {\gamma}/ d \log {\rho}$. When $\gamma$  is small, the gas is ``soft" with respect to
compression by the flow, and large densities may be attained compared to
gases with larger $\gamma$. If $\gamma$ is
significantly smaller than unity, an approximately power-law ${\rm f}(\rho) $ is predicted, while for
$\gamma$  close to unity ${\rm f}(\rho)$ should have a  nearly lognormal form (Scalo et al.
1998, Passot \& Vazquez-Semadeni 1998, Nordlund \& Padoan 1999).  To the extent
that the SFR is some increasing function of the integral of ${\rm f}(\rho) $ at large
densities, we expect the SFR to be much larger in the power-law (small-$\gamma$) case
compared to the lognormal (large-$\gamma$) case, due to the presence of a longer tail at large densities in the former case (cf. Fig. 10 of Scalo et. al. 1998). The situation is more complicated when the role of magnetic field
fluctuations is considered ( Vazquez-Semadeni \& Passot 2000), but protogalaxies are expected to have negligible magnetic fields (Ruzmaikin et al. 1988) and hence we neglect this effect here.

The present
paper tries to estimate the effective softness, as measured by the effective barotropic index ${\gamma} $, for a turbulent protogalaxy
whose cooling is dominated by ${\rm H_2}$ (and HD at the lowest temperatures). 
Our approach does not
include the details of ${\rm H_2}$ + HD chemistry and UV feedback, but instead
evaluates ${\gamma} $ for a galaxy in which ${\rm H_2 + HD}$ cooling dominates, with
the ${\rm H_2}$ and HD fractions given as parameters.  As will be seen, the
behavior of $\gamma $ is largely independent of the ${\rm H_2}$ fraction as long as it is small.  Our goal is only to map out the regions of the
temperature-density space in which ${\rm H_2 + HD}$ cooling yields ``soft" ( $ \gamma 
< 1$) or ``hard" ( $\gamma \ga 1$) behavior, which suffices to estimate the
relative fraction of the gas driven to large densities.
       
 We emphasize that our view of the SFR is fundamentally different
from the model proposed by Norman \& Spaans (1997) and Spaans \& Norman
(1997); see also Spaans \& Carollo (1997).  These authors presented
detailed comprehensive models for many of the physical processes as a
function of redshift, but ignored the potential effects of turbulent
compressions.  For this reason they were led to assume that efficient
star formation will only occur when sufficient metal production had
occurred so that fine-structure cooling would induce a phase transition
via thermal instability, at redshifts $\sim 1-2$.  Besides, in view of the possibility
that turbulence effectively suppresses the thermal instability phase
transition (Vazquez-Semadeni, Gazol \& Scalo 2000), this picture neglects
the possibility that high density condensations can be formed by
turbulence even in a pure ${\rm H_2}$ galaxy, if the value of ${\gamma} $ is
sufficiently less than unity. Our view of turbulent cloud and star formation is similar to the picture evvisioned by Elmegreen (1993) and Padoan (1995), except for the sensitivity to $\gamma$.

	 A study complementary to the present work has been given by
Spaans \& Silk (2000), who calculated $\gamma$ for conditions expected in
quasi-static self-gravitating molecular (i.e. shielded) clouds with
metallicities ${\rm Z}> 0.01{\rm Z}_{\odot}$.  Their calculations include a detailed
treatment of the chemistry and radiative transfer, various coolants
not considered in the Z=0 cases treated here, the thermal coupling
between gas and dust, and heating by the infrared Cosmic Background Radiation, which is
important for redshifts between about 10 and 40 if the metallicity is
large enough to allow significant dust abundances.  They also discuss
many implications of the resulting $\gamma$ for stellar masses,
high-redshift star formation, and starburst galaxies, all in the
context of the ability of self-gravitating objects to collapse.  In
contrast, the present work is concerned with the ${\rm Z} <0.01{\rm Z}_{\odot}$ case,
parameterizes the chemistry in terms of assumed ${\rm H}_2$ and HD abundances,
neglects dust grains (probably justified for these small
metallicities), and is focused on the maximum densities, and hence
star formation rates, that can be attained in a {\it turbulent} 
galaxy or cloud.  Primarily because of the larger assumed metal
abundances, the values of $\gamma$ derived by Spaans and Silk and their
dependence on density is somewhat different than found here for a
pure ${\rm H_2 + HD}$ gas.

In section 2, we show explicitly how $\gamma$ is determined by the logarithimic derivatives of the heating and cooling rates. Numerical evaluation of these derivatives and $\gamma$ for a gas containing ${\rm H_2}$ and HD are presented in section 3. Our interpretation of these results and their applicability to galaxies are discussed in section 4.

\section {The Effective Barotropic Index}

There are several physical processes by which turbulent interactions affect the density field, and these are mediated by $\gamma$. A specific process
for density amplification would be the supersonic interaction of two
velocity streams or clouds, which will form a dense slab which may be
subject to gravitational instability (Elmegreen 1993) and Vishniac's
(1994) nonlinear bending mode instability (Blondin et al. 1996, 
Klein \& Woods 1998). Vazquez-Semadeni, Passot \& Pouquet (1996) pointed out that, while the
density jump behind a shock resulting from such an interaction is expected to be of order ${\rm Ma^2}$ (Ma = Mach
number) for $\gamma   = 1$ (isothermal flows), the density jump approaches ${\rm e^{{Ma}^2}}$
as $\gamma= 0$. Also the minimum Mach number
required to induce gravitational instability in a given region is a
strong function of ($1 - \gamma$), as shown in various contexts by Hunter \&
Fleck (1982), Tohline et al. (1987) among others. The importance of the effective barotropic index ($\gamma$) of the ISM of
galaxies was explored by numerical simulations by Vazquez-Semadeni et al. (1996) in the context of supersonic
turbulence.

  For a large range of conditions relevant for the interstellar gas in
galaxies it is true that, except for regions immediately behind shock
fronts, the dynamical timescale of the gas motions in a region of size L
and characteristic velocity v, L/v, is much larger than the cooling time
due to microscopic processes.  This inequality holds for a large number
of environments, including diffuse HI and molecular gas in our own and
other galaxies, including protogalaxies (depending on the scale of
interest).  The inequality implies that the heating and cooling terms in
the fluid energy equation will easily adjust to balance each other on the
timescale of the dynamical flows, i.e. the gas can be considered to be in
thermal equilibrium, as first emphasized by 
 Vazquez-Semadeni et al. (1996).  The velocity and density fields are controlled by
the momentum and continuity equations, while the temperature (and hence
pressure) are slaved to the relatively slowly-varying density field
through the thermal equilibrium condition, which determines a unique
temperature (and pressure) for any given density.  In effect, the gas
behaves as a barotropic fluid with pressure given by $P = {\rho}^{\gamma}$
     where
${\gamma}$    is determined by the thermal equilibrium condition.

  A general expression for $\gamma$ can be derived from the thermal
equilibrium condition

\begin{equation}                       
\Lambda(\rho,{\rm T}) = \Gamma (\rho,{\rm T})
\end{equation}

\noindent where $\Lambda$ and $\Gamma$ are the cooling and heating rates per unit volume. 
The index $\gamma$ is defined by

\begin{equation}
\gamma \equiv \frac{{\rm d logP}}{{\rm d log}{\rho}} = 1+ \frac{{\rm d log T}}{{\rm d log} \rho}
\end{equation}

\noindent where the perfect gas equation of state was used in the last step and we
neglect variations in the mean molecular weight.
    Defining the function ${\rm F}(\rho, {\rm T}) = \Gamma(\rho ,{\rm T}) -
 \Lambda(\rho, {\rm T}) = 0$   (by eq.1)
, we have after implicit differentiation,

\begin{equation}
\frac{{\rm d log T}}{{\rm d log} \rho} =  \frac{\frac{\partial {\rm log} \Gamma}{\partial{\rm log} \rho}- \frac{\partial {\rm log} \Lambda }{\partial {\rm log} \rho }} {\frac{\partial {\rm log} \Lambda } {\partial {\rm log T} }- \frac{ \partial {\rm log} \Gamma}{\partial {\rm log} T}} 
\end{equation}                       

\noindent $\gamma$ then follows from eq. 2.  Our goal is to evaluate the derivatives
of the cooling function for cases in which the dominant coolants are  ${\rm H_2}$ and HD.
In the present work we use the detailed calculations of  ${\rm H_2}$ cooling by
LeBourlet, Pineau des Forets, \& Flower (1999) and Flower et al. (1999), which include collisions
with H, ${\rm H_2}$, and He, and use quantum mechanical cross-sections that
supersede previous work.  The disagreement in $\gamma$  that would be
obtained using previous work can be seen by examining the slopes of the
cooling curves presented in Figure A1 of Galli \& Palla (1998).

For the heating rate we take the parameterized
form $\Gamma \propto {\rho}^a{\rm T}^b$. The simplest case is for optically thin heating
by UV photons or heating by cosmic rays, in which case a = 1  and b = 0
to good approximation.  Since in true primordial protogalaxies with no
stars it is difficult to imagine such heating, we are assuming that some
low level of star-forming activity has already occurred. 
   It is instructive to notice that if the cooling rate is similarly
parameterized as $ \Lambda = {\rho}^c{\rm T}^d$, then

\begin{equation}
\gamma = 1+(1-c)/d 
\end{equation}

\noindent If c=1 (for a = 1 and b = 0) , this immediately shows that for densities large enough for
collisional deexcitation to be important in controlling the level
populations, so that $\Lambda  \sim {\rho}$ (instead of ${\rho}^2$ at lower densities), the
effective barotropic index will be close to unity (isothermal), and hence
the turbulence will be relatively ``hard."  The common assertion,
supported marginally by observations, that dense molecular gas without
internal star formation should be nearly isothermal is due to the fact
that the densities in in these regions exceed the critical density for CO
collisional deexcitation, not because the CO cooling is particularly
efficient, as is often stated.  In fact the range of densities for this
condition to hold for CO is relatively small, as pointed out by Scalo et
al. (1998).  The same sort of effect will be seen below for ${\rm H_2}$+HD.

  We next examine the behavior of the logarithmic derivatives of the
${\rm H_2}$ and HD cooling rates, and evaluate the turbulent
compressibility for the case of a simple UV or cosmic ray dominated
heating rate.

\section{ Results}
From the expression for $\gamma $ derived in the previous section we see that it is determined by the derivatives of the cooling function. Thus, in order to obtain the correct compressibility it is important to determine the proper cooling function (and its derivatives) for the primordial gas. Fortunately, making a theoretical estimate of the cooling function in the primordial universe is simpler than the corresponding problem in the present day universe, where there exists a large number of coolants.

Since Saslaw \& Zipoy(1967) first discussed the importance of the ${\rm H_2}$ 
molecule in cosmology it has been established that main coolant in the early 
universe below a temperature of $10^4$ K is the ${\rm H_2}$ molecule (see 
Galli\& Palla 1998 for a review of astrochemistry of the early universe, 
and Stancil, Lepp and Dalgarno 1996 on the possible importance of other 
molecules, such as HD and LiH). The calculation of the ${\rm H_2}$ abundance is considerably complicated by the non-equilibrium nature of the chemical kinetics (Shapiro \& Kang, 1987), which elevates the  ${\rm H_2}$  abundance compared
to its equlibrium concentration. The detailed chemistry of ${\rm H_2}$ formation is not treated here, and
we take the ${\rm H_2}$/H ratio as a parameter.  Estimates of the 
${\rm H_2}$/H fraction by
Tegmark et al. (1997) and Ferrara (1998) suggest an ${\rm H_2}$ fraction of
 $10^{-3}$
for protogalaxies.  However a detailed consideration of the chemical
 pathways for ${\rm H_2}$ formation by Galli \& Palla (1998) gives for average cosmic gas density ${\rm H_2/H}
 \sim 10^{-6}$ for redshifts less than about 100. At high redshifts ($z \geq 100$), ${\rm H_2}$ formation is inhibited 
even in overdense regions because the required intermediaries ${\rm H_2}^+$ 
and ${\rm H}^-$ are dissociated by cosmic microwave background(CMB) photons. 
The ${\rm H_2}$ abundance can be as high as 
$10^{-3}$ inside collapsed clouds (Abel \& Haiman 1997, Abel et al. 2000; see section 4.1 for discussion). 
 We will show that the dependence of $\gamma$ on ${\rm H_2}$ 
fraction is very
 small for
 ${\rm H_2}$ fractions within three orders of magnitude of the adopted value 
of $10^{-6}$, because $\gamma$    only depends on the logarithmic derivatives 
of the cooling function with
 respect to temperature and density, not the absolute value of the cooling
 function itself.  This result becomes invalid if the ${\rm H_2 / H}$ fraction
 is a
 strong function of temperature or density, which would need to be
 included in the logarithmic derivatives. We neglect this effect here, 
since they require detailed chemistry calculation. We have calculated $\gamma$ for ${\rm H_2}$ abundances of $10^{-6}$ and 
$10^{-3}$ here.

  Le Bourlot et al. (1999) have recently calculated the cooling functions for a gas containing ${\rm H_2}$ molecules, which are collisionally excited
 by H, He, and ${\rm H_2}$. They have  computed the ${\rm H_2}$ cooling rates per molecule from a detailed computation of non-LTE level populations 
and quantum-mechanical collisional cross-sections, assuming the gas to be optically thin to these transitions. They have provided a dataset of
calculated cooling values in a wide temperature $(10^2 -10^4
  {\rm K})$, density $(1-10^8 {\rm cm}^{-3})$, ${\rm n_H/n_{H_2}}$ $ (10^{-8}$ to $10^6)$, and ${\rm n_{ortho}/n_{para}}$  (0.1 to 3) range that  we have adopted for calculating cooling values for our purpose. The results are insensitive to $ {\rm n}_{\rm ortho}/{\rm n}_{\rm para}$ which we take as unity.


 Although the HD abundance is much less than ${\rm H_2}$, the presence of a non-zero permanent dipole moment and smaller rotational constant make it a potentially more important coolant at lower temperatures. The question of HD abundance in the post-recombination era is also well-studied (see Puy et al 1993, Tegmark et al. 1997 and references therein). We have adopted values from Galli \& Palla (1998) which give this ratio to be $ {\rm [HD/H_2]} = 1.1 \times 10^{-3}$.

The HD cooling function is taken from Flower et al (1999), who provide a
routine for calculating cooling values due to HD rotational transitions within
the vibrational ground state of HD, collisionally excited by H, ${\rm H_2}$, and He, again assuming optically thin transitions.

The ${\rm H_2}$ and HD cooling rates have been combined to estimate the net cooling rates for a wide  temperature $(10^2 -10^4 {\rm K})$ and density
 $(1-10^8 {\rm cm}^{-3})$ range. These are shown in Figure 1.  Since Le Bourlot et al. have provided cooling rates only for values of temperature 
and density which are too widely spaced for accurate  calculation of derivatives by direct differentiation, we first needed to produce a local 
polynomial fit for the cooling function to avoid ``jittery" derivatives that will result from their linear interpolation program. 
From the requirement of continuous second-order derivatives we used a cubic spline interpolation of the cooling function. 

\begin{figure}
\epsfxsize=8truecm
\epsffile{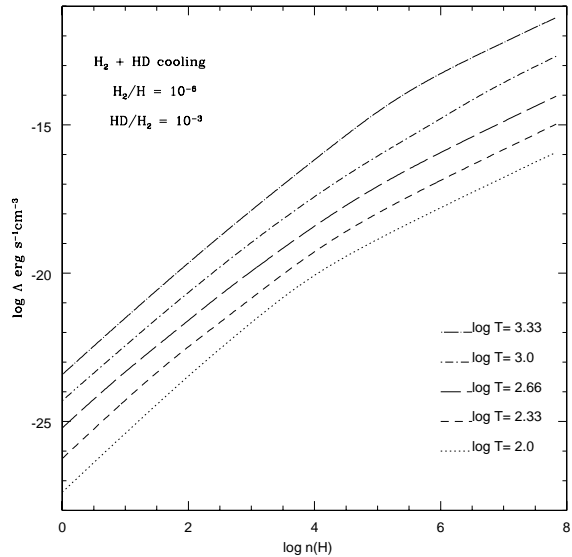}
\caption{  Logarithm of total (${\rm H}_2$ + HD) cooling function per unit volume as a function of logarithm of density for five different
values of temperature ; ${\rm H}_2$/H = $10^{-6}$ and HD/${\rm H}_2$ = $1.1 \times 10^{-3}$.}
\end{figure}

In the  adopted parameterized heating rate $\Gamma \propto {\rho}^a{\rm T}^b$, the constant of proportionality is arbitrary. We do not adopt any 
particular value for this constant of proportionality as the cosmic ray and/or UV fluxes  in protogalaxies are highly uncertain, depending upon 
(among other things) the unknown star formation rate. We incorporate this arbitrary scaling of the magnitude of the heating rate by letting the 
temperature be an independent variable, not coupled to the density by the thermal equillibrium condition (which we use only to calculate 
logarithimic derivatives). Thus the labelling of curves in our plots by temperature is essentially a labelling by the amplitude of the heating rate. 
 The temperature range in actual protogalaxies may be inferred by comparison with the hydrodynamic simulations of Abel et al. (2000),
 which suggest that T = 200-800K may be appropriate. A brief discussion of $\gamma$ in the case where adiabatic heating dominates is given in sec. 4.1 below.

\begin{figure}
\epsfxsize=16truecm
\epsfysize= 16truecm
\epsffile{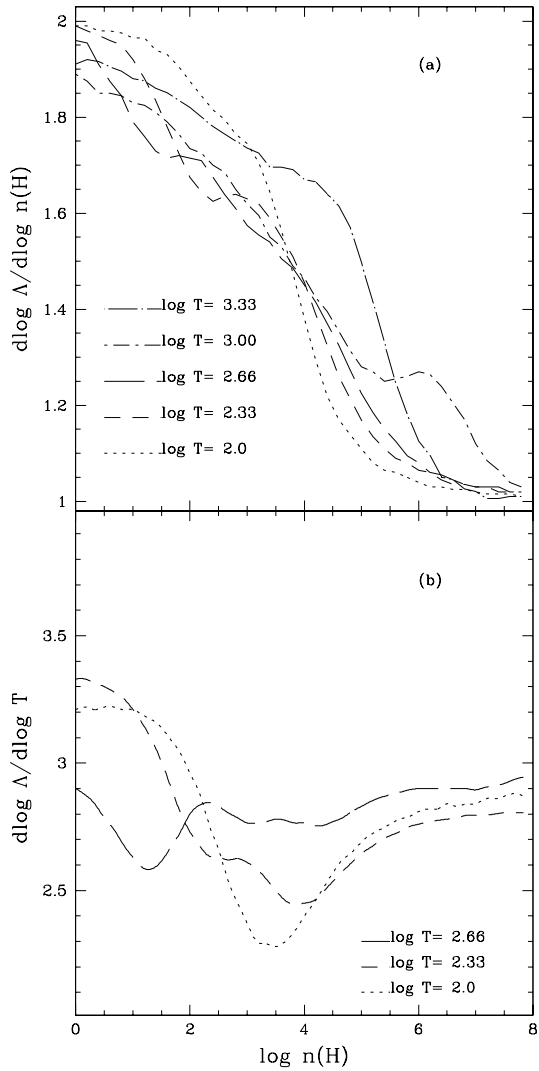}
\caption{Logarithmic derivative of total cooling function with respect to density (a) and temperature (b) as  a function of the logarithm of density for different temperatures; ${\rm H}_2$ and HD ratios same as in Figure 1.}

\end{figure}

Figures 2(a) and 2(b) show the logarithmic derivative of the ${\rm H}_2 +$ HD cooling function with respect to logarithmic density and temperature.
 The derivative with respect to logarithmic density has a value close to 2 at low densities and 1 at high densities as anticipated. 
The transition of cooling function from a quadratic dependence on density to a linear dependence takes place around the critical density above which 
collisional deexciation dominates radiative decay as discussed, for example in Spitzer(1978). The derivatives with respect to temperature
 do not show appreciable variation with density at higher temperature and remain more or less constant at a value close to 3.
 This general behaviour can be shown using the analytic prescription for the ${\rm H}_2$ cooling function given by Tegmark et al. ( 1997) based on
 Hollenbach and McKee (1979).

\begin{figure}
\epsfxsize=16truecm
\epsfysize = 16truecm
\epsffile{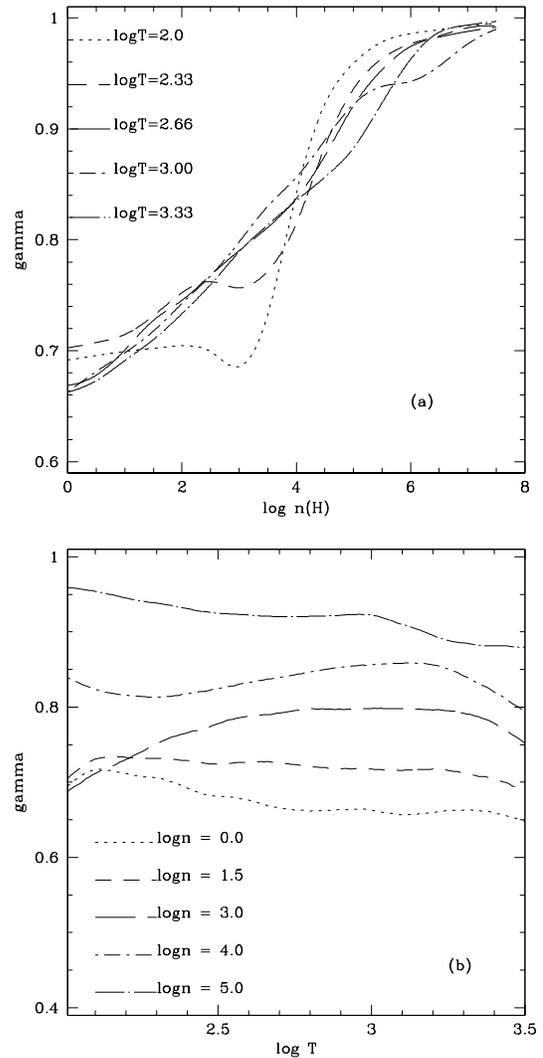}
\caption{ Barotropic index ($\gamma$)  as a function of the logarithm of density(a) and temperature(b); ${\rm H}_2$ and HD ratios same as in Figure 1.}
\end{figure}

Plots of turbulent  compressibilty ($\gamma$) for various temperatures and H number density for ${\rm [H_2/H]}=10^{-6}$ and ${\rm [HD/H_2]}=1.1 
\times 10^{-3}$ are shown in Figures 3(a) and 3(b) respectively. From Figure 3(a) it is noticed that  at the lowest temperatures $\gamma$ values shoot up 
near the critical density of HD (see section 4.3) as collisional deexcitation takes over from radiative decay. On the other hand, at high 
temperatures  $\gamma$ varies quite smoothly with density. We attribute this to the fact that at higher temperatures the higher energy states 
are populated, and there is no single critical density characteristic of a two-level system, as pointed out by Le Bourlot et al. (1999) and others. 
Overall, between temperatures of 100 to 2000 K, above which ${\rm H_2}$ molecules begin to dissociate, $\gamma$ remains within a range of 0.7 to 1. 
It takes a large value of 1 at large densities because cooling rates become proportional to number density. With decreasing density,  $\gamma$ 
decreases as the cooling function becomes proportional to $n^2$ in the low density limit. Fig. 3(b) shows that there is not much dependence of $\gamma$ on temperature for a given density. 

\begin{figure}
\epsfxsize=16truecm
\epsfysize= 16truecm
\epsffile{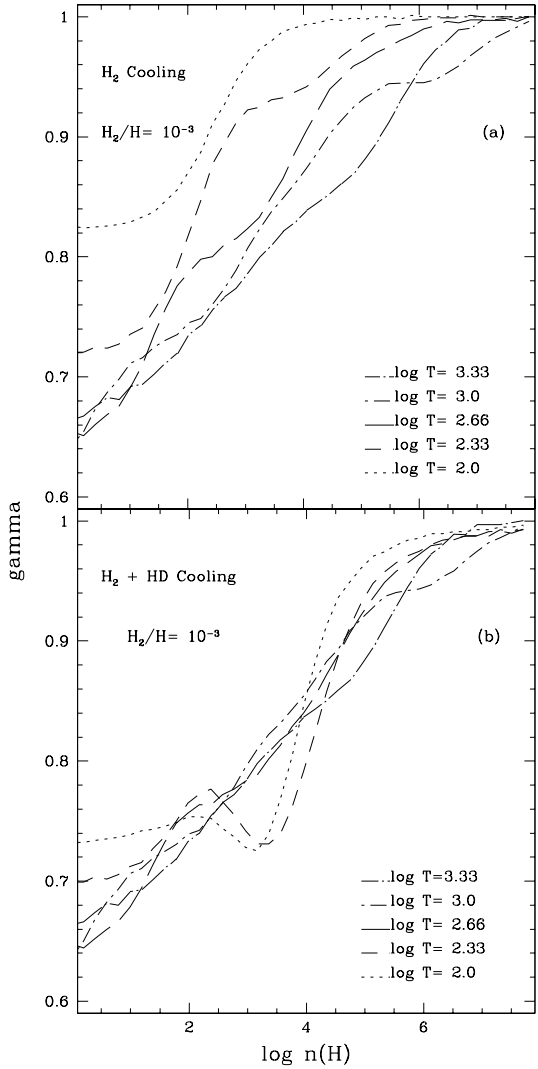}
\caption{Barotropic index ($\gamma$) as a function of the logarithm of density, when  the cooling is due to ${\rm H}_2$ molecule only (a), and due to both ${\rm H}_2$ and HD molecule (b);  ${\rm H}_2$/H = $10^{-3}$ and HD/${\rm H}_2$ = $1.1 \times 10^{-3}$.}
\end{figure}

In  Figures 4(a) and 4(b) we  show  how  $\gamma$ varies with density when the ${\rm H}_2$/H ratio is different ($10^{-3}$). 
The HD/${\rm H}_2$ ratio is assumed to be same as earlier. A comparison of Figure 4(b) with Figure 3(a) shows there is little difference in 
$\gamma$ due to change in ${\rm H}_2$/H ratio within three orders of magnitude. Figure 4(a) shows how $\gamma$ varies with density when HD is absent.
 A comparison of Figure 4(a) with 4(b) shows that at small densities and low temperatures {\it {the gas is much harder in absence of HD} }. This implies that
 HD cooling is ``softer" at low densities than ${\rm H}_2$ cooling.
At higher temperatures the two plots look almost identical because the contribution of HD to the total cooling function becomes insignificant in those temperatures.

\section{Discussion}

\subsection {Density Range}

 Because our derived values for the turbulent compressibility
depend sensitively on whether the density is above or below the
effective critical density for collisional deexcitation, it is
important to understand the density ranges expected in protogalactic
objects.  Although no certain conclusion can be reached, suggestive
results can be found in the recent highly resolved simulations of
Abel, Bryan, \& Norman (2000).  Abel et al. found that primordial
molecular clouds with masses $ \sim 10^5 {\rm M}_{\odot}$ are formed at the
intersections of filaments at redshifts around 40, with average ${\rm H}_2$
fraction less than about $10^{-4}$ and particle densities much less than
$1 {\rm cm}^{-3}$.  The density at the center of the most massive clump during
the period Z= 35 to 23 is about 0.3 to 3 ${\rm cm}^{-3}$, with T rising from
$ \sim 200 {\rm K}$  to  $\sim 800 {\rm K}$  due to adiabatic heating, and f(${\rm H}_2$) increasing from
about $10^{-5}$ to $10^{-4}$ during this period.  After redshift $\sim 23$ the
density increases rapidly, reaching $10^4 {\rm cm}^{-3}$.  In the densest core
the cooling time is comparable to the free fall time, so our
assumption of fast cooling is not valid.

 For these reasons our calculations are applicable only to the
GMC-like structures which have smaller densities.  The largest
densities are presumably due to the fact that the low-density gas is
highly compressible, as we have explained in terms of the basic
cooling physics here, while we expect the evolution of the dense
cores, even when aided by self-gravity, to be relatively ``hard",
since at these large densities the ${\rm H}_2$ is nearly in LTE and the value
of $\gamma$ should be large.  The value of $\gamma$ probably remains large
during the contraction of the cores if they are able to
gravitationally collapse (Spaans \& Silk 2000). 

 This can be seen as follows. If adiabatic (compressional) heating at the free-fall rate dominates, then using $ \nabla . {\vec u} = - d \log {\rho}/dt \sim 1/{ {\tau}_{ff}} $  in the internal energy equation shows that the heating rate scales as $ \rho ^{3/2} $ (using $ {\tau}_{ff} \sim \rho^{-1/2}$). Using this heating rate and reading off the logarithmic derivative of the cooling rate from Fig. 2, equations 2 and 3 show that at densities $ n = ( 10^2, 10^4, 10^6)$, $\gamma \approx (0.8, 1.0, 1.3) $. The latter value of $\gamma$ holds approximately for all densities greater than $ 10^6 cm^{-3}$. Thus the gas becomes increasingly ``hard" with increasing density. Only at the lowest densities is the gas moderately ``soft". This is similar to what we found for diffuse heating, except that the values of $\gamma$ are larger and the densities required for ``soft" behavior are smaller in the compressional heating case. Notice that for $ n \ga 10^6$ the value of $\gamma$ is sufficiently close to the critical value of $4/3$ that classical fragmentation should be prevented, i.e. the thermal Jeans mass cannot decrease with increasing density. A similar result was found by Spaans \& Silk (2000) who consider gas-grain heating rather than adiabatic compressional heating. This implies that the mass associated with densities around $10^6 {\rm cm}^{-3}$ will be the maximum mass of Pop III objects. However the result is tentative, since our assumption of thermal balance may not be valid for collapsing objects.

\subsection{ Metallicity and Redshift}

Our results are only relevant for protogalaxies whose
metallicity is so small that ${\rm H}_2$ and HD cooling dominates.  
Based on
the detailed calculations of Norman \& Spaans (1997) and Spaans \&
Norman (1997), the transition metallicity is Z $ \sim 0.01{\rm Z}_{\odot}$.  At larger
Z cooling by fine structure atomic and ionic lines dominates for the
densities and temperatures of interest.  We adopt this as the
critical metallicity here.

The question of the redshift at which this critical Z is reached in a
cosmic-averaged sense cannot be answered at present.   Metals and dust have been observed in galaxies with redshifts of at least 5 (Armus et al. 1998). Observations of
Fe and, to a lesser extent, Zn in high-redshift Ly $\alpha$ clouds
(Prochaska \& Wolfe 2000) indicates that the column density-weighted
Fe abundance relative to the solar system is about 0.025 for
redshifts between 1 and 4, and is remarkably constant.  There is a
large range in Fe abundance at given redshift, but only a small
fraction of Ly $ \alpha $ clouds have Fe abundances as small as the
estimated critical value, suggesting that the ${\rm H}_2$ cooling-dominated
phase terminates at redshifts less than at least 4 on average.  Pettini et al.
(1997) estimate the average metallicity in damped Ly $ \alpha $ systems to
be $ \sim 0.05 {\rm Z}_{\odot}$ at Z=3.  For the lower column density Ly $ \alpha $ forest
clouds, carbon abundances (Tytler et al. 1995, Songaila \& Cowie 1996
and references therein) are approximately 0.01 times solar, but the
overall metallicity (dominated by oxygen) may be larger.  An extreme
lower limit for the metallicity of the Ly $ \alpha $ forest at redshift 3
is $0.001 {\rm Z}_{\odot}$ (Songaila 1997).
        Observations of both emission and absorption lines intrinsic
to QSOs give solar or higher metallicities out to redshifts of at
least 4 (see Hamann 1998 and references therein).  How these results
relate to metallicities in the bulk of galactic gas is currently
unknown, but do show that the present results are only applicable to
galactic nuclei at much larger redshifts.
        Based on studies of the intergalactic medium and elliptical
galaxies in clusters, Renzini (1998) has argued that the mean
metallicity of the universe at ${\rm z} \sim 3$ is about $0.1 {\rm Z}_{\odot}$.

Theoretically, the problem is extremely complicated, since the metal
enrichment rate depends on both the star formation rate, which is
essentially a parameter in all models and is only very weakly
constrained by observations, the IMF, which is unknown, the ejection
of metals from galaxies by supernova-energized outflows, and other
factors.  The redshift corresponding to Z= $ 0.01{\rm Z}_{\odot}$ in the Norman \&
Spaans (1997) models was about 3.  A recent calculation of the
evolution of the cosmic metallicity by Pei, Fall, \& Hauser (1999),
assuming an IMF lower mass limit of $0.1{\rm M}_{\odot}$, shows that metallicities
as small as $0.01{\rm Z}_{\odot}$ occur only for redshifts greater than 5,
consistent with the limit from the Ly $ \alpha$ observations.  As pointed
out by Padoan et al. (1999) and others, if the IMF is strongly
weighted toward high-mass stars, metal enrichment might only take
about $ 10^7$ yr because the metal yield (mass of metals produced
relative to total mass incorporated into a generation of stars) will
be large.  A plot of yields as a function of IMF lower mass cutoff is
given in Scalo (1990).
        Another way to look at the problem is to assume a redshift
for the formation of the first star-forming objects and timescale for
metal enrichment.  For example, the simulations of Abel et al. (2000)
produce GMC-mass clouds at redshift ${\rm Z} \sim 40$, in rough agreement with
some previous non-simulation estimates (e.g. Tegmark et al. 1997).
If star formation proceeds with a ``normal" IMF, the duration of metal
enrichment may be about $10^8$ years, corresponding to a critical
metallicity of $0.01{\rm Z}_{\odot}$ at redshift 15 (assuming $\Lambda =0, {\rm q} = 1/2,
{\rm H}_0 =  75$ cosmology).  If the protogalaxy IMF is extremely top-heavy,
then the enrichment may only take about $10^7$ yr, corresponding to a
redshift of about 34  for the same cosmology.

We conclude that if the IMF in galaxies is not extremely
top-heavy, then our calculations of $\gamma$ for a gas dominated by
${\rm H}_2$ +HD cooling may be relevant, in a cosmic-averaged sense, from the
epoch of galaxy formation to redshift about 4-5, with great
uncertainty. In the ``worst case" scenario in which star
formation was rapid and the IMF was heavily weighted to massive
stars, the duration of metal enrichment $\sim 10^7$ yr implies that the
${\rm H}_2$-dominated phase covered the narrow redshift range from around 40  to
30.  However the cosmic average may not be very
relevant.  One general conclusion that follows from the observational
considerations above is that the universe was chemically very
inhomogeneous at redshifts up to $\sim 5$.

\subsection { The critical density}

\begin{figure}
\epsfxsize=16truecm
\epsfysize= 8truecm
\epsffile{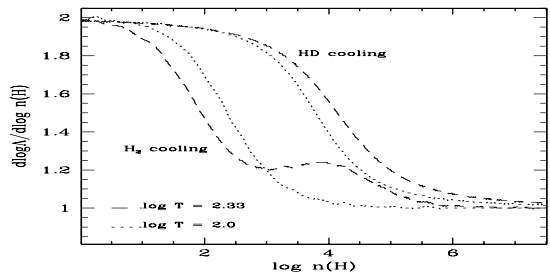}
\caption{Logarthmic derivatives of the cooling functions due to HD and ${\rm H}_2$ separately as
 functions  of the logarithm of density at T= 100K and 200K . ${\rm H}_2$ and HD ratios same as
in Figure 4.}
\end{figure}

 To use these results to predict SFR in protogalaxies we need to have predictions about the critical value of $\gamma$ below which the density pdf 
develops a power-law tail. This tail behavior occurs at $\gamma$ values as large as 0.7 to 0.8 in simulations by Passot and Vazquez-Semadeni (1998) 
and Scalo et al.(1998). So depending upon the critical value of $\gamma$, densities from several hundreds to several thousands should be commonly 
attained by turbulent compressions. If gravitational collapse can commence at such densities, then the SFR in protogalaxies should be appreciably 
larger than that has been previously speculated ( e.g. Spaans \& Norman 1997, Norman \& Spaans 1997). On the other hand, $\gamma$ approaches a value 
of unity above densities $ \sim 10^5$. So if such densities or higher are required for commencement of gravitational collapse then the SFR should be 
low in protogalaxies. In any case we note that the ${\rm H}_2 +$ HD gas can only attain a lowest value of ${\gamma}$ of 0.7, which is still considerably
``harder'' than for fine structure cooling of atomic gas. Thus the conclusion of Norman \& Spaans (1997) and Spaans \& Norman (1997) that the SFR should
peak when the fine structure cooling commences may be valid, but because of the differences in turbulent compressibility, not because of thermal instability.

The most interesting result is that at low temperatures the gas becomes abruptly hard  above the critical density for collisional deexcitation for HD, 
implying that structure formation at those temperatures is unlikely to occur above the critical density.
We conclude that the structures that form by turbulent interactions at temperatures below a few hundred degrees can be characterized by an upper 
density limit which is the critical density for the equality of collisional and radiative deexcitation, since the value of $\gamma$ approaches unity 
for larger densities. This result is similar to the situation for CO-dominated cooling in contemporary molecular clouds where $\gamma$ is small 
below the critical density for collisional de-excitation (Scalo et al. 1998). This result implies that realistic density distributions for 
protogalactic or present day large metalicity galaxies can only be obtained from simulations if the  details of cooling function are accounted for; 
isothermal simulations will clearly underestimate the densities of clouds formed by turbulence.

 We predict that the
distribution of gas densities formed by turbulent interactions should
have a peak corresponding to the critical density for HD collisional
deexcitation, which is $ \sim 10^4 {\rm cm}^{-3}$. It is important to note that if HD cooling is absent the critical density would be determined
by  ${\rm H}_2$ , which has a critical density of $\sim 10^2$, as seen from Fig. 5. It is interesting that densities larger by a factor of hundred
can be attained by turbulence in a universe like ours, with relatively large D abundance, compared to a situation in which D abundance is much
smaller.

Galaxies with larger metal
abundances should have a peak at densities corresponding either to
fine structure cooling (if molecules cannot form) or carbon monoxide
(dominant coolant for molecular gas at moderate densities) cooling. The
critical densities for collisional deexcitation in case of fine structure cooling would be
$ \sim 10^3
 {\rm cm}^{-3}$ if CII is the dominant coolant and $\sim 10^5 {\rm cm}^{-3}$ if OI is the main species. 
The critical density when CO is the dominant coolant is roughly $ \sim 10^3 {\rm cm}^{-3}$ depending on the temperature (see Goldsmith \& Langer 1978). If the medium is optically thick to its own radiation, then radiative trapping can also reduce the density dependence of the cooling rate; however a proper treatment must include radiative line transfer, and 
there is currently no viable model for line transfer in a turbulent medium.  Existing calculations that use escape probabilities show the 
sensitivity of the radiative trapping critical density to the adopted velocity gradient parameter (Goldsmith \& Langer 1978, Qaiyum \& Ansari 1987) 
for CO.

\noindent We appreciate detailed comments by Enrique Vazquez-Semadeni.

\label { lastpage}
\end{document}